\documentclass[prb,twocolumn,amsfonts]{revtex4}
\usepackage{graphicx}
\usepackage{amsmath}
\usepackage{amsfonts}
\usepackage[dvips]{epsfig}
\usepackage{bm}
\usepackage[active]{srcltx}
\usepackage{epsf}
\usepackage{subfigure}

\def\ie{{\it {i.e.}\ }}

\begin{document}
\author{Samuel Marcovitch, Yakir Aharonov, Tirza Kaufferr and Benni Reznik }
\title{Combined Electric and Magnetic Aharonov-Bohm Effects}
\affiliation{ School of Physics and Astronomy,
Raymond and Beverly Sackler Faculty of Exact Sciences,
Tel-Aviv University, Tel-Aviv 69978, Israel.}

\date{\today}
\begin{abstract}

It is well-known that the electric and magnetic Aharonov-Bohm effects may be formally
described on equal footing using the four-vector potential in a relativistic framework.
We propose an illustrative manifestation of both effects in a single configuration, in which the
specific path of the charged particle determines the weight of the electric and magnetic acquired relative phases.
The phases can be distinctively obtained in the Coulomb gauge.
The scheme manifests the pedagogical lesson that though each of the relative phases is gauge-dependent
their sum is gauge-invariant.

\end{abstract}
\maketitle
%_____________________________________________________________________________
%\section{Introduction}
The Aharonov-Bohm (AB) \cite{ab} effect is traditionally separated into two different effects:
the electric AB effect and the more familiar magnetic one.
These may be formally described on equal
footing using the four-vector potential in a relativistic framework, where the acquired relative phase is
$\oint A^\mu d x_\mu$.
However, a physical manifestation of their simultaneous existence may in fact be achieved with
a simple specific configuration in space-time.
In this paper we set forth such a manifestation, which enables vivid tangible understanding
of the relationship between the two effects.

Both the electric and the magnetic AB effects
realize the significance of electromagnetic potentials on charged particles which travel through
regions of space that are not simply connected.
Even if such particles encounter no electromagnetic fields in their paths,
they may acquire a relative phase resulting in a shift in the interference pattern.
In the electric effect, a charged particle's wave packet passes through a
non-simply connected region,
where it is split in two;
each of the two wave packets encounters a different
scalar potential that can only depend on time.
Therefore, no electric forces act on the particle.
The relative phase acquired is
$\oint\phi dt$ in Gaussian units where we take $\hbar=c=e=1$.

\begin{figure}[ht]
\center{
\includegraphics[width=2.5in]{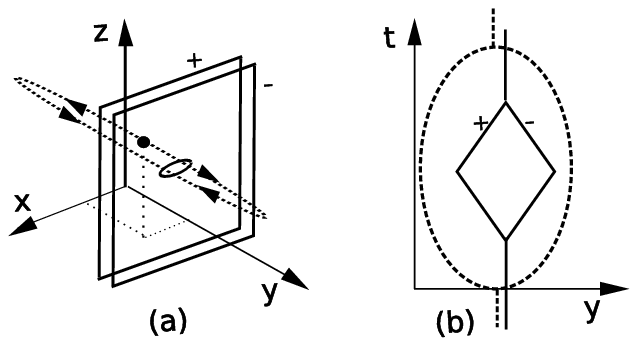}
\caption{A charged particle's wave-packet (the dashed line) is split in the $y$ direction.
Each wave packet travels to a different side of a large capacitor, whose plates are zero distance apart.
Then the two plates are moved apart and then together again in the $y$ direction
(the solid line).
Afterwards the wave packets interfere.
Note that one of the wave packets travels back and forth through a little hole in the capacitor.
(a) space dependence, (b) time dependence.
}
\label{el2}
}
\end{figure}

This effect may be physically realized in the following way:
A charged particle's wave-packet
starts on one side of a capacitor and is split
%in the $y$ direction
so that one of the wave packets travels through the capacitor
while the other goes in the opposite direction,
as shown in figure \ref{el2}a.
The capacitor is initially uncharged.
We assume there is a little hole in the capacitor through which the wave packet travels undisturbed.
Then the capacitor is charged and discharged, and
only afterwards the wave packets return together and interfere
(so that one of the wave packets travels again through the hole).
We assume that the wave packets are well apart from the boundaries of the capacitor (or that the capacitor plates are infinite along the $x$ and $z$ axes) and that the hole is small.
Therefore, each of the wave packets travels through a region of zero electric field.
Actually, instead of charging and discharging the capacitor, we can just move the plates of a charged capacitor: initially, two oppositely charged plates are zero distance apart so that they totally cancel each other. Then the capacitor plates are moved apart and then together again in the $y$ direction.
This setup is illustrated in figure \ref{el2}b.

In the magnetic AB effect, a charged particle's wave packet interferes around a confined magnetic flux,
acquiring a relative phase of
$\oint \boldsymbol{A} \cdot d \boldsymbol{x}=\oint \boldsymbol{B} \cdot d \boldsymbol{S}$,
where we take the magnetic susceptibility and the electric permittivity to be 1.
This effect may be physically realized using a long solenoid, as shown in figure \ref{mag2}:
\begin{figure}[ht]
\center{
\includegraphics[width=2.5in]{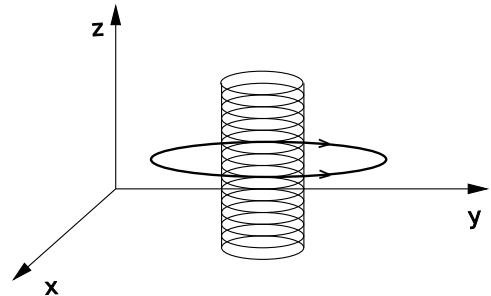}
\caption{A charged particle's wave packet interferes around a long solenoid.}
\label{mag2}
}
\end{figure}

Our scheme is based on constructing a subspace in space-time in which the four-vector potential is non trivial.
Though the sources that yield the suggested potential depend on time,
we arrange the setup to be non-radiating
so that both the electric and magnetic fields vanish everywhere
outside the source.
When a charged particle's wave packet encounters such a potential, it acquires an AB phase both from the
electric and magnetic effects, which depend on the particle's specific path in space-time
$ \theta=\oint\boldsymbol{A}\cdot d \boldsymbol{l}+\oint\phi dt$.
By expressing the potential in the Coulomb gauge we see which part of the phase is electric and which is magnetic.
This way we see how the electric and magnetic relative phases depend on different possible paths.

\begin{figure}[ht]
\center{
\includegraphics[width=2.2in]{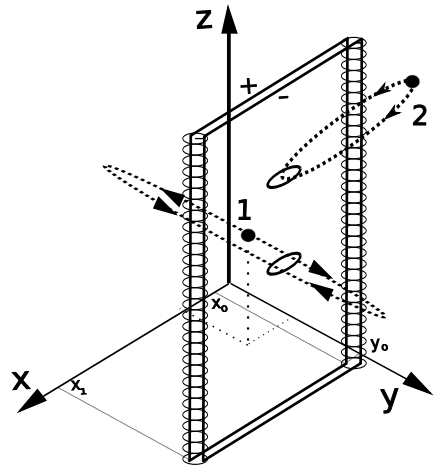}
\caption{Two capacitor plates are infinite in the $z$ direction,
finite in the $x$ direction and zero distance apart in the $y$ direction.
In addition, two infinitesimally thin and infinitely long solenoids
are located on the edges of the capacitor.
Paths $1$ and $2$ describe interfering wave packets of a charged particle.
In path $1$ (the "electric path") the wave packet is split in the $y$ direction. One of the wave packets travels back and forth through a little hole in the capacitor.
In path $2$ (the "magnetic path") the wave packet is split in the $y$ direction and travels in the $x$ direction. Then the two wave packets interfere through a second hole in the capacitor.
}
\label{fig42}}
\end{figure}

The setup that combines both effects is described in figure \ref{fig42}.
%Its time dependency is described in figure \ref{fig4}.
Assume that the two capacitor plates are infinite in the $z$ direction
%, finite in the $x$ direction
and zero distance apart in the $y$ direction.
In the $x$ direction the plates are finite where the edges are located at $x_0$ and $x_1$ and we define $L=x_1-x_0$.
In addition, two infinitesimally thin and infinitely long solenoids
(fluxons) are located on the edges of the capacitor (along the $z$ axis).

\begin{figure}[ht]
\center{
\includegraphics[width=3.0in]{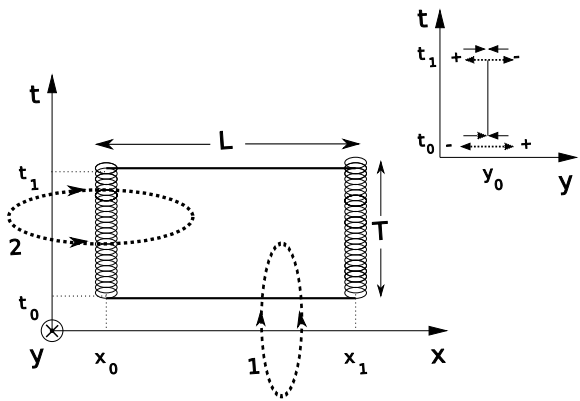}}
\caption{At $t=t_0$ the two plates are instantaneously moved apart and then together again.
Simultaneously, steady opposite currents are sent through the two solenoids beginning at $t=t_0$ and ending at $t=t_1$.
At $t=t_1$ the plates are again instantaneously moved apart and then together, but in the opposite direction (as shown in the small graph).
%and the fluxons are simultaneously annihilated.
A charged particle, whose interfering wave packet is described in the "electric path"
(path $1$), acquires both an electric and magnetic phases.
A charged particle whose interfering wave packet encircles one of the fluxons, as described in the "magnetic path" (path $2$), acquires a magnetic phase alone.
\label{fig4}}
\end{figure}

The time-dependence of the setup is illustrated in figure \ref{fig4}.
At $t=t_0$, the plates are instantaneously moved apart and then together again.
At $t=t_1$, we apply the same procedure again:
%on the capacitor's plates:
the plates are instantaneously moved apart and then together again,
but in the opposite direction
(as shown in the small graph in figure \ref{fig4}).
In order to cancel out the resulting electromagnetic wave fronts,
we simultaneously send steady opposite currents through the two solenoids during the time-interval $[0,T]$,
where $T=t_1-t_0$.
%simultaneously with the annihilation of the fluxons.
We note that this choice of sudden changes in the charge and current densities is not essential and is used to simplify the calculations, as shall be explained later.
As we now show, the suggested configuration describes a non-radiating source, which creates a non-simply connected topology in space-time.
A charged particle that enters this finite volume in the $xt$ plane acquires a phase from both
the electric and magnetic effects.


First, consider a particle moving along path $1$ of figures \ref{fig42} and \ref{fig4},
which we regard as the "electric path":
at $t<t_0$, its wave packet is located at $(x,y,z)$
where $x_0<x<x_1$, $y<y_0$ and $z$ is arbitrary (and will be omitted from now on). The wave packet then splits in the $y$ direction, so that at $t=t_0$, the two wave packets are located on both sides of the capacitor; one is located at $(x,y_1)$, where $y_1>y_0$ and the other at $(x,y_2)$, where $y_2<y_0$. Then at $t_0<t<t_1$ the wave packets interfere at $(x,y)$.
%The path is illustrated as path $1$ .
As we shall later see the relative phase between the wave packets in this path is acquired from both the electric and magnetic effects.

Second, consider a particle moving along path $2$ of figures \ref{fig42} and \ref{fig4},
which we regard as the "magnetic path".
At $t_0<t<t_1$, its wave packet is located at $(x,y_0)$
where $x<x_0$. The wave packet then splits in the $y$ direction and travels in the $x$ direction. Then the wave packets interfere, so that they close a loop around one of the fluxons well within the time interval $[t_0,t_1]$.
%This path is illustrated as path $2$ of figures \ref{fig42} and \ref{fig4}.
Here again, the wave packets travel through a little hole in the capacitor, when the plates are zero distance apart.
%Therefore, the wave packets are undisturbed.
As we shall later see the relative phase in this path is acquired solely from the magnetic effect.
\\
\begin{figure}[ht]
\center{
\includegraphics[width=1.6in]{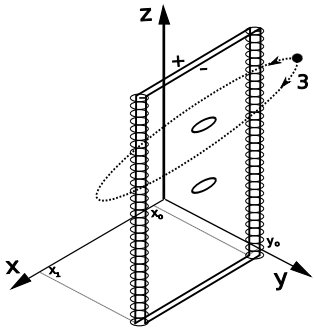}}
\caption{Path $3$: A charged particle wave packet interferes around the whole setup.
\label{fig43}}
\end{figure}

One may ask what would be the relative phase acquired by wave packets that interfere without crossing the capacitor, as shown in path $3$ of figure \ref{fig43}.
It seems that the relative phase should be only an electric one, since the two magnetic phases acquired by each of the opposite fluxons cancel each other.
However, in this setup no relative phase is acquired at all.
This is explained in the following by choosing a specific gauge in which the four-vector potential vanishes along the path of the wave packets.

We continue by explicitly showing that this setup is indeed non-radiating and derive the relative phases.
For convenience, let us take $x_0=t_0=y_0=0$.
We define the following singular vector potential:
\begin{equation}
\label{A}
\boldsymbol{A}=\pi\big[\theta(t)-\theta(t-T)\big]\big[\theta(x)-\theta(x-L)\big]\delta(y)\hat{\boldsymbol{y}},\quad \phi=0,
\end{equation}
where
\begin{displaymath}
\theta(x)=\left\{ \begin{array}{ll}
0, & x<0\\
1/2, & x=0\\
1, & x>0
\end{array} \right.
\end{displaymath}
is the Heaviside step function.
Note that even though there are sources in the suggested setup, we choose to fix $\phi$ to zero,
so that the gauge is fixed as the temporal gauge.
$\boldsymbol{A}$ can be visualized as confined to the rectangle shown in figure \ref{fig4}, and pointing in the $y$ direction.
We have imposed delta function $\delta(y)$ thus making the potential singular purely for simplicity of calculations; it is in no way an essential requirement \cite{rem}.

We then see that both the electric and magnetic fields
vanish everywhere outside the source given by $\boldsymbol{A}$:
\begin{equation}
\begin{split}
&\boldsymbol{E}=-\frac{\partial\boldsymbol{A}}{\partial t}=
-\pi\big[\delta(t)-\delta(t\!-\!T)\big]\big[\theta(x)-\theta(x\!-\!L)\big]\delta(y)\hat{\boldsymbol{ y}},
\\&\boldsymbol{B}=\!\frac{\partial \!A_y}{\partial \!x}\!-\!\frac{\partial A_x}{\partial y}\!=\!
\pi\!\big[\theta(t)\!-\!\theta(t\!-\!T)\big[\!\delta(x)\!-\!\delta(x\!-\!L)\big]\delta(y)\hat{\boldsymbol{z}}.
\end{split}
\end{equation}
The electric field appears momentarily at $t=0$ and $t=T$ and the magnetic field is constant during $[0,T]$.
The sources, \ie charge and current densities, yielding the vector potential
(\ref{A}) can be easily found using
%the continuity equation,
the differential form of Gauss' Law and the potential form of Maxwell's equations
(corresponding to the current gauge):
\begin{eqnarray}
\label{he}
%&&\boldsymbol{\nabla}\cdot \boldsymbol{j}+\partial\rho/\partial t=0,
%\nonumber
&&\boldsymbol{\nabla}\cdot \boldsymbol{E}=4\pi\rho,
\nonumber
\\&&\nabla^2\boldsymbol{A}-\frac{\partial^2\boldsymbol{A}}{\partial t^2}-
\boldsymbol{\nabla}(\boldsymbol{\nabla}\cdot\boldsymbol{A})=-4\pi\boldsymbol{j}.
\end{eqnarray}

The resulting charge density is
\begin{equation}
\label{dens}
\rho=-\frac{1}{4}\big[\delta(t)-\delta(t-T)\big]\big[\theta(x)-\theta(x-L)\big]\delta'(y).
\end{equation}
We see that indeed the charge density corresponds to
capacitor plates that are instantaneously moved apart and then together again
in the $y$ direction both at $t=0$ and $t=T$
(from the derivative of the delta function).

We separate the resulting current density $\boldsymbol{j}$ into two
parts $\boldsymbol{j}=\boldsymbol{j}_c+\boldsymbol{j}_s$, where $\boldsymbol{j}_c$
corresponds to the capacitor plates' momentary movements at $t=0$ and $t=T$ and
$\boldsymbol{j}_s$
corresponds to two infinitesimally thin solenoids
that are turned on at $t=0$ and turned off at $t=T$, located on the edges of the capacitor:
%(so that $\boldsymbol{\nabla}\cdot\boldsymbol{j}_s=0$):
\begin{equation}
\label{js}
\begin{split}
\boldsymbol{j}_c=&\frac{1}{4}\big[\delta'(t)-\delta'(t-T)\big]\big[\theta(x)-\theta(x-L)\big]\delta(y))\hat {\boldsymbol{ y}},
\\ \boldsymbol{j}_s=&
\frac{1}{4}\big[\theta(t)-\theta(t-T)\big]
\boldsymbol{\nabla}\times\Big\{\big[\!\delta(x)\!-\!\delta(x\!-\!L)\big]\delta(y)\hat{\boldsymbol{z}}\big]\Big\}.
\end{split}
\end{equation}
One can verify that the continuity equation
$$\boldsymbol{\nabla}\cdot \boldsymbol{j}+\partial\rho/\partial t=0$$ holds.

Let us now separate the acquired phase into the electric phase and the magnetic phase.
As (\ref{A}) shows, the relative phase acquired when a charged particle's wave packet crosses
$y=0$ is
$$\theta=\oint A^\mu d x_\mu=\pi\big[\theta(t)-\theta(t-T)\big]\big[\theta(x)-\theta(x-L)\big],$$
which equals $\pi$ if the particle crosses the discussed topology (figure \ref{fig4}).
This phase may originate from the electric effect induced by the charges
as well as the magnetic one induced by the solenoidal currents.

In order to explicitly obtain the electric and
magnetic parts of the phase we transform to the Coulomb gauge \cite{remark}.
In this gauge the scalar potential depends only on the charge density
and the vector potential depends only on the transverse current density $\boldsymbol{j}_t$
for which $\boldsymbol{\nabla}\cdot\boldsymbol{j}_t=0$: \cite{jl}
\begin{equation}
\label{c}
\begin{split}
&\nabla^2\phi=-4\pi\rho,
\\& \nabla^2\boldsymbol{A}-\partial^2\boldsymbol{A}/\partial t^2=-4\pi\boldsymbol{j}_t.
\end{split}
\end{equation}
Therefore the electric and magnetic phases are clearly distinct in the Coulomb gauge.

Gauge transformation of the four-vector potential
\begin{equation}
\begin{split}
&\boldsymbol{A}\rightarrow\boldsymbol{A'}=\boldsymbol{A}+\boldsymbol{\nabla}\Lambda,
\\&\phi\rightarrow\phi'=-\partial\Lambda/\partial t
\end{split}
\end{equation}
to the Coulomb gauge, where $\boldsymbol{\nabla}\cdot\boldsymbol{A'}=0$, requires
solving the Poisson equation, $\nabla^2\Lambda=-4\pi C$, in two dimensions
(as the setup is invariant under replacements along the $z$ axis),
where
\begin{equation}
C=\frac{1}{4\pi}\boldsymbol{\nabla}\cdot\boldsymbol{A}=
\frac{1}{4}\big[\theta(t)-\theta(t-T)\big]\big[\theta(x)-\theta(x-L)\big]\delta'(y).
\end{equation}
The Green's function of Poisson's equation in two dimensions, $\nabla^2\Lambda=4\pi\delta(x-x_0)\delta(y-y_0)$,
is \cite{morse}
$$G(x,y|x_0,y_0)=Re\big[-2\ln(\omega-\omega_0)\big],$$ where $\omega\equiv x+iy$.
Therefore,
\begin{equation}
\Lambda =\int C(x',y',t') \ \ln\big[(x-x')^2+(y-y')^2\big]dx'dy'.
\end{equation}
Twice integrating by parts and canceling out the boundary terms we get,
\begin{equation}
\Lambda =\big[\theta(t)-\theta(t-T)\big]F(x,y),
\end{equation}
where
\begin{equation}
\label{fxy}
F(x,y)=\frac{1}{2}\big[\arctan(\frac{x}{y})- \arctan(\frac{x-L}{y})\big].
\end{equation}
The corresponding potentials are then:
\begin{equation}
\label{cpot}
\begin{split}
\phi'&=-\frac{1}{2}\big[\delta(t)\!-\!\delta(\!t-\!T)\big]\!
\left[\!\arctan\!\left(\frac{x}{y}\!\right)\!-\!\arctan\!\left(\!\frac{x-L}{y}\right)\!\right],
%\\A_x'&=\big[\theta(t)\!-\theta(\!t-T)\big]\frac{\partial F(x,y)}{\partial x}
\\A_x'&=\frac{1}{2}\big[\theta(t)\!-\theta(\!t-T)\big]\!
\left[\frac{y}{x^2+y^2}-\frac{y}{(x-L)^2+y^2}\right],
%\\A_y'&=\big[\theta(t)\!-\theta(\!t-T)\big]\frac{\partial F(x,y)}{\partial y}
\\A_y'&=-\frac{1}{2}\big[\theta(t)\!-\theta(\!t-T)\big]\!
\left[\frac{x}{x^2+y^2}-\frac{x-L}{(x-L)^2+y^2}\right].
\end{split}
%\nonumber
\end{equation}
Note that $A_y'$ is non-singular now since the singular term $A_y$ cancels out,
as can be verified from the first equation in (\ref{delta}).
%&\phi'=-\frac{1}{2}\!\big[\delta(t)-\delta(\!t\!-\!T\big)],
%\\&A_x'=\big(\theta(t)-\theta(t-T\big))\frac{L(L-2x)y}{2(x^2+y^2)\big((L-x)^2+y^2\big)}.
In addition, one can verify (\ref{c}) by taking into account that
%\cite{jl}
\begin{equation}
\label{delta}
\begin{split}
&\underset{x\to 0}{\lim}\frac{\partial}{\partial y}\arctan\left(\frac{x}{y}\right)=
-\pi\big[\theta(x)-\theta(-x)\big]\delta(y),
\\ &\underset{x\to 0}{\lim}\frac{\partial}{\partial x}\left(\frac{x}{x^2+y^2}\right)=2\pi\delta(x)\delta(y),
\\ &\underset{x\to 0}{\lim}\frac{\partial}{\partial y}\left(\frac{x}{x^2+y^2}\right)=0,
\end{split}
\end{equation}
where we have used $\theta(x)+\theta(-x)=1$ and the definition
\begin{equation}
\underset{x \to 0}{\lim} \frac{1}{\pi}\frac{x}{x^2+y^2}=\big[\theta(x)-\theta(-x)\big]\delta(y).
\end{equation}
%All other derivatives are non singular.

In order to manifest the path-dependence of the electric and magnetic phases,
let us first discuss wave packets that interfere in the "electric path",
corresponding to path $1$ of figure \ref{fig4}.
If the charged particle's wave packets are located at $(x,d)$, $(x,-d)$ at $t=0$ and interfere before $t=T$
%where %$0<d<<L$ and $0<x<L$ and $d>0$,
then
the obtained electric and magnetic phases are:
\begin{equation}
\label{tetae}
\begin{split}
\theta_e&=\oint\phi' dt=-\big[F(x,-d)-F(x,d)\big]
\\&=\arctan\left(\frac{x}{d}\right)-\arctan\left(\frac{x-L}{d}\right),
\\\theta_m&=\oint\boldsymbol{A'}\cdot d \boldsymbol{l}=\int_{d}^{-d} \left(\frac{\partial F(x,y)}{\partial y}+A_y\right)dy
%\\&=\arctan\left(\frac{d}{x}\right)-\arctan\left(\frac{d}{x-L}\right).
\\&=-\theta_e+\pi\big[\theta(x)-\theta(x-L)\big].
\end{split}
\end{equation}
Obviously, the sum of the electric and the magnetic phases equals $\pi$ if $0< x < L$ and vanishes otherwise.
%\begin{equation}
%\begin{split}
%\theta&=\theta_e+\theta_m=\oint\boldsymbol{A'}\cdot d \boldsymbol{l}+\oint\phi' dt
%\\&=\int_{y_1}^{y_2}\!A'_y(x,y,t)dy+\int_{t_1}^{t_2}\!\phi'(x,y,t)dt\!=\!\pi,
%\end{split}
%\end{equation}
%where $y_1 y_2\!<0$, $t_1 t_2\!<0$ and $\max(t_1,t_2)<T$.

\begin{figure}[ht]
\center{
\includegraphics[width=3in]{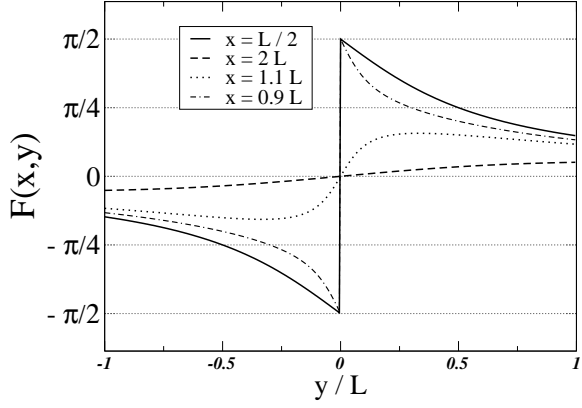}
\caption{$F(x,y)$. Space dependence of $\phi'(x,y,t)$, the scalar potential in the Coulomb gauge,
is given as a function of $y$ for various values of $x$.
If the particle's wave function consists of two wave packets, which are located at $(x,y_1)$ and $(x,y_2)$ at $t=0$
and interfere before $t=T$, then the electric phase is given by
$F(x,y_1)- F(x,y_2)$ (for $y_1>y_2$).
For $0< x < L$, $F(x,y)$ has singularity in $y=0$, for which the sum of the electric and magnetic phases is non-trivial.
}
\label{phi}
}
\end{figure}

%The space dependent term of the scalar potential $\phi'$, namely
$F(x,y)$ (\ref{fxy})
is illustrated in figure \ref{phi} as a function of $y$ for various values of $x$ at $t=0$.
From (\ref{tetae}) it can be immediately seen that
if the distance $2d$ between the wave packets in the $y$
direction at $t=0$ is small compared to L
(that is $0<d<<L$) and $0<x<L$,
the electric phase equals $\pi$ and thus the magnetic one vanishes.
As $d$ increases, the electric phase becomes less dominant,
while the magnetic phase becomes more dominant as their sum still equals $\pi$.
% and the magnetic phase becomes more dominant, yet their sum still equals $\pi$:
The decrease in the electric phase can be physically explained by the fact that the capacitor is now finite.
%Since the capacitor is not infinite, the electric phase decreases.
Note that if $x(t=0)<0$, $x(t=0)>L$ or both wave packets are located in the same side of the capacitor at $t=0$,
the electric and magnetic phases cancel.
Furthermore, if the wave packets interfere only after $t=T$, the total phase is zero.
%$$ \int_{y_1}^{y_2}dy A'_y(x,y) + \phi'(x,y_1)-\phi'(x,y_2)=0. $$

Secondly, let us examine wave packets that interfere in the "magnetic path" corresponding to path $2$ of figure \ref{fig4}.
It can easily be checked that wave packets that interfere around $(0,0)$ or $(L,0)$ (not both)
well within $[0,T]$
acquire only a magnetic phase of $\pi$.
From the definition of the vector potential (\ref{A}), it can be immediately seen that no relative phase is acquired in path $3$ (figure \ref{fig43}). Here again the magnetic and electric phases cancel.
%it can be immediately seen that both phases vanish.
\\
\\

\begin{figure}[ht]
\center{
\includegraphics[width=2.5in]{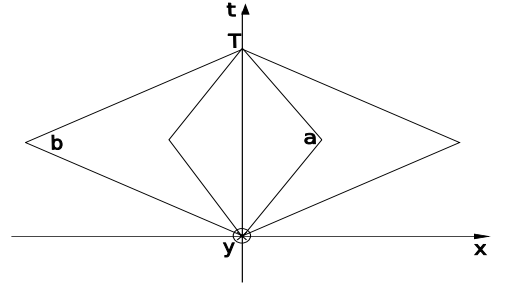}
\caption{a. Time-like setup. The vector potential points towards the $y$ direction and is confined inside the rhombus.
It describes two opposite magnetic dipoles that travel through the vector potential boundaries and by two electric dipoles
that travel in the same track, where the electric part is weaker. b. Space-like setup, describing the same scheme but where the electric
dipoles dominate.}
\label{fig5}
}
\end{figure}

Let us now generalize the above scheme for any suggested singular gauge, manifesting
a symmetry that gives a deeper insight into the proposed scheme.
For simplicity, let space be two-dimensional.
Then in the previous scheme the solenoids in figure \ref{fig4} represent two opposite magnetic dipoles at rest
and the horizontal lines represent two wires of the capacitor that are momentarily opened and closed.
We can transform the rectangular potential in figure \ref{fig4} to a boosted one in which the magnetic dipoles travel on its boundaries.
Assume for example that the potential is transformed to: 
%Let us use this in order to explore the charge and current densities of
%the set-up given by the following potential:
%We can boost each of the two magnetic dipoles in opposite directions along the $x$ axis.
%perform a Lorentz transformation to (\ref{A}) so that at $t=0$ the  and
%boost the system and
%This yields a
%combination of magnetic and electric dipoles.
%For example, consider the time-like setup described by the following vector-potential
\begin{equation}
\label{AA}
\begin{split}
&\boldsymbol{A}\!=\!\pi\!\theta(\!vt\!-\!x)\theta(\!vt\!+\!x)\theta\!\big[\!\!-\!v(\!t\!-\!T\!)\!+\!x\big]\!\theta\!\big[\!\!-v(\!t\!-\!T\!)\!-\!x\big]\!\delta(\!y\!)\hat{\boldsymbol{y}},
\\&\phi=0,
%\nonumber
\end{split}
\end{equation}
which corresponds to the rhombus in figure \ref{fig5}.
%(where the product of the step functions produces the interior of the rhombus).
First, if $v=0$ we reduce to a case of just two opposite magnetic dipoles at rest at $x=0$.
Next, for $0<v<1$, we get the time-like setup which is visualized in figure \ref{fig5}a.
In this setup the magnetic dipoles are boosted, therefore the densities correspond to both magnetic and electric dipoles.
Using the same considerations as in (\ref{he})
%and that the product of the delta function and the theta function
%equals the delta function or zero,
we obtain charge and current densities
which correspond to two opposite magnetic dipoles and two electric dipoles that travel on the boundary of the vector potential.
The magnetic dipoles travel in opposite directions along the $x$ axis.
The electric dipoles start traveling in opposite directions along the $x$ axis
with positive sign during the interval [0,T/2]
and then are flipped to negative sign during [T/2,T].
Since $v<1$ the magnetic part is more dominant than the electric one:
\begin{equation}
\label{dipol}
\begin{split}
&\rho=-\frac{v}{2}\big[f_+(t,x)\!-f_-(t,x)\big]\delta'(y),
\\&\boldsymbol{j}_s=\frac{1}{2}\boldsymbol{\nabla}\times\Big\{\big[g_+(t,x)-g_-(t,x)\big]\delta(y)\boldsymbol{\hat z}\Big\},
\end{split}
%\nonumber
\end{equation}
where $f_{\pm}(t,x)$ define the paths for the electric dipoles, and
$g_{\pm}(t,x)$ define the paths for the magnetic dipoles:
\begin{equation}
\begin{split}
f_+(t,x)=&\big[\delta(vt- x)+\delta(vt+ x)\big]
\big[\theta(t)-\theta(t-T/2)\big]
\\ &\big[\theta(x+vT/2)-\theta(x-vT/2)\big],
\\f_-(t,x)=&\Big\{\delta\big[-(v(t-T)- x\big]+\delta\big[-v(t-T)+ x\big]\Big\}
\\&\big[\theta(t\!-\!T/2)\!-\!\theta\!(t\!-\!T)\!\big]\!
\big[\!\theta(x\!+\!vT/2)\!-\!\theta(x\!-\!vT/2)\!\big]\!,
\\g_+(t,x)=&\big[\delta(vt- x)+\delta(-v(t-T)- x)\big]
\\&\big[\theta(t)-\theta(t-T)\big]\!
\big[\theta(x)-\theta(x-vT/2)\big],
\\g_-(t,x)=&\big[\delta(vt+ x)+\delta(-v(t-T)+ x)\big]
\\&\big[\theta(t)-\theta(t-T)\big]\!
\big[\theta\left(x+vT/2)-\theta(x)\right].
\end{split}
\end{equation}

Is there also a meaning to a space-like setup as illustrated in figure \ref{fig5}b, with $v>1$?
As a matter of fact we can illustrate the two horizontal lines in figure \ref{fig4} as two electric
dipoles located at the center of the wire at $t=0$ that travel with infinite velocity in opposite directions.
(Take $v\to\infty$ in (\ref{AA}). Of course, no real superluminal velocity is assumed.)
In such a system no magnetic dipoles appear.
However, if we ``boost'' the electric dipoles with a finite velocity greater than $1$ we obtain a
potential corresponding to figure \ref{fig5}b,
resulting in both electric and magnetic dipoles, while now the electric part dominates,
as can be seen from (\ref{dipol}).
%In the light-like setup the weights of the electric and magnetic dipoles are the same.

\begin{figure}[ht]
\center{
\includegraphics[width=3in]{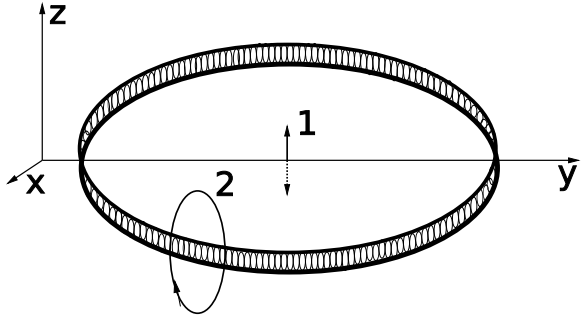}
\caption{Poloidal current flows inside an infinitesimal toroidal solenoid during $[0,T]$.
The capacitor's plates are simultaneously moved apart and then together again
at $t=0$ and at $t=T$ (in opposite directions).
A charged particle's wave packet that travels in paths 1 or 2 can acquire both
electric and magnetic phases.}
\label{fig3}
}
\end{figure}

As a final remark, we note that the combination of the AB effects can
%The setup suggested in  requires a capacitor that is infinite along the $z$ axis.
also be manifested using finite capacitors, specifically circular capacitors,
as illustrated in figure \ref{fig3}.
Here the configuration is similar to (\ref{A}).
We instantaneously move the plates apart and then together again.
At the same time a constant current $\boldsymbol{j}_s$
flows through an infinitesimally thin toroidal solenoid that encircles the
capacitor just at the boundaries of the capacitor.
The current is poloidal (loops around the torus).
Then after a period of time $\Delta T$ we instantaneously move the capacitor plates apart and then together again,
but in the opposite directions, and simultaneously send an opposite current through the toroidal solenoid,
so that the currents cancel each other.
%This setup is illustrated in figure \ref{fig3}.
Analogously to (\ref{A}), we choose
\begin{equation}
\label{a3d}
\boldsymbol{A}=\pi\big[\theta(t)-\theta(t-T)\big]\theta(L-r)\delta(z)\hat{\boldsymbol{z}},\quad \phi=0,
\end{equation}
expressed in cylindrical coordinates.
From (\ref{he}) we get the predicted charge and current densities:
\begin{equation}
\begin{split}
&\rho=-\frac{1}{4}\big[\delta(t)-\delta(t-T)\big] \theta(\!  L  -r)\delta'(z),
\\ &\boldsymbol{j}_s=\frac{1}{4}\big[\theta(t)-\theta(t-T)\big]\boldsymbol{\nabla}\times
\big[\delta(L-r)\delta(z)\boldsymbol{\hat{\phi}}\big].
%\Big\{\big[& \frac{\delta(L-r)}{r}+\delta'(L-r)\big]\delta(z)\hat{\boldsymbol{ z}}-
%\\& \delta(L-r)\delta'(z)\hat{\boldsymbol{ r}}\Big\}.
\end{split}
\end{equation}
The electric and magnetic phases
are extracted by transforming the vector potential (\ref{a3d}) to the Coulomb gauge.
A somewhat similar non-radiating setup
has also been proposed by Afanasiev \cite{afan},
where the current was taken to be linearly dependent on time and the capacitor was static.
Such a setup was proposed there as a time-dependent AB effect.
In this respect, the proposed scheme may suggest non-classic information transfer applications using
time-dependent AB effects.
While in the original magnetic AB effect
only one topological bit of information is encoded in the relative phase,
in the time-dependent setup many bits of information can be encoded
in the topology without sending any classical traces such as electromagnetic fields into the rest of the world.

In conclusion, we would like to emphasize the following point.
Gauge invariance as a classical property is manifested in quantum mechanics by the acquirement of a relative phase
when the wave function passes through a non-simply-connected region.
The electric and magnetic AB effects are distinctively manifested in the Coulomb gauge.
However, in any configuration, the total relative phase, electric plus magnetic, is gauge invariant.

%____________________________________________________________________________________________________________________________
\section{Acknowledgments}
We thank J. Kupferman, S. Popescu, A. Casher and M. Marcovitch for helpful discussions.
This work has been supported by the European Commission under the
Integrated Project Qubit Applications (QAP) funded by the IST
directorate as Contract Number 015848 and by the Israel Science Foundation grant number 784/06.
%____________________________________________________________________________________________________________________________


\begin{thebibliography}{99}


\bibitem{ab} Y. Aharonov and D. Bohm, Phys. Rev., 115, 485 (1959).

\bibitem{rem}
The same effect can be obtained by choosing a smooth non-singular vector potential
$\boldsymbol{A}({\bf x},t)$
that is confined in both space and time.
Outside the confined region the potential vanishes.
Within a certain part of the confined region, it is constant.
In the intermediate area it smoothly decays from a constant value to zero, and it is  only in this region that the electric and magnetic fields do not vanish.
Such a potential imposes charge and current densities different from (\ref{dens},\ref{js}), yet the setup is still non-radiating.
The same combined electric and magnetic Aharonov-Bohm effects are obtained, if the two interfering wave packets do not enter (in spacetime) the intermediate regions, where $\bf E$ and $\bf B$ are
non-trivial.
We have used the delta (and step) function(s) to simplify the calculations, specifically those of
the charge and current densities.
%Alternatively, one can replace $\delta(y)$ in (\ref{A}) with a Lorentzian function $x/\pi(x^2+y^2)$ and go on with the calculations with the non-singular potential with the same physics.

\bibitem{remark} When treating the sources quantum mechanically, the Coulomb gauge is essential, since
it describes the back-reaction correctly.
Y. Aharonov and J. Anandan, Phys. Let. A, 160, 493 (1991).

\bibitem{jl} Note that $\boldsymbol{j}=\boldsymbol{j}_l+\boldsymbol{j}_t$, where $\boldsymbol{j}_l$ is the longitudinal current
for which $\boldsymbol{\nabla}\times\boldsymbol{j}_l=0$. This standard separation of $\boldsymbol{j}$ does not coincide with
$\boldsymbol{j}_c$, $\boldsymbol{j}_s$ in (\ref{js}) as can be verified from (\ref{c},\ref{cpot}).


\bibitem{morse} P.M Morse and H. Feshbach, Methods of Theoretical Physics (McGraw-Hill, New-York, 1953), p-1175.

\bibitem{afan} G.N. Afanasiev and Yu P. Stepanovsky, J. Phys. A: Math, Gen. 28, 4546 (1995);
G.N. Afanasiev, Topological Effects in Quantum Mechanics (205-206), Kluwer Academic Publishers (1999).









\end{thebibliography}
\end{document}